\documentclass[12pt]{article}

\usepackage{graphicx}
\usepackage{pgfplots}
\usepackage{booktabs}
\usepackage{amsmath}
\usepackage{algorithm}
\usepackage{algorithmic}
\usepackage{amssymb}
\usepackage{amsfonts}
\usepackage{latexsym}
\usepackage{color}
\usepackage{subfigure}
\usepackage{latexsym}
\usepackage{graphicx,epstopdf}
\usepackage{subfigure}
\usepackage{authblk}
\usepackage{caption}
\usepackage{dsfont}

\newcommand{\be}{\begin{equation}}
\newcommand{\en}{\end{equation}}
\newcommand{\bea}{\begin{eqnarray}}
\newcommand{\ena}{\end{eqnarray}}
\newcommand{\beano}{\begin{eqnarray*}}
\newcommand{\enano}{\end{eqnarray*}}
\newcommand{\bee}{\begin{enumerate}}
\newcommand{\ene}{\end{enumerate}}

\newcommand{\Lc}{{\cal L}}
\newcommand{\D}{{\cal D}}

\newcommand{\1}{1 \!\! 1}
\newtheorem{thm}{Theorem}

\newtheorem{prop}[thm]{Proposition}

\begin{document}

\title{Reply to Comment on "A no-go result for the quantum damped
	harmonic oscillator"}

\author{F. Bagarello$^{1,2}$, F. Gargano$^{1}$, F. Roccati$^3$\\

\small{
$^1$DEIM -  Universit\`a di Palermo,
Viale delle Scienze, I--90128  Palermo, Italy,}\\
\small{ $^2$I.N.F.N -  Sezione di Napoli,}\\
\small{
$^3$Dipartimento di Fisica e Chimica Emilio Segr\`e, Universit\`a degli Studi di Palermo, 
via Archirafi 36, I-90123 Palermo, Italy.}\\
\small{\emph{Email addresses:}\\
fabio.bagarello@unipa.it,
francesco.gargano@unipa.it,
federico.roccati@unipa.it}
}

\date{}
\maketitle
\begin{abstract}
In a recent paper, \cite{deguchi}, Deguchi and Fujiwara claim that our results in \cite{BGR} are wrong, and  compute what they claim is the square integrable vacuum of their annihilation operators. In this brief note, we show that their vacuum is indeed not a vacuum, and we try to explain what is behind their mistake. We also consider a very simple example clarifying the core of the problem. 
\end{abstract}


\section{Introduction}\label{sec:intro}

The problem of quantizing dissipative systems, and the damped harmonic oscillator (DHO) is not an easy task. There exist different approaches of very different kind. In this short note we only concentrate on the Bateman's approach, which is based on the use of a virtual {\em amplified} oscillator (AHO), coupled with the DHO. In \cite{nakano} the authors claimed they can quantize the full system by using ladder operators which look formally quite close to what one of us introduced some years ago under the name of $\D$ pseudo-bosons, see \cite{baginbagbook} for a review. The main idea in \cite{nakano} is that the Bateman Hamiltonian $H$ can be written in a diagonal form, and that its eigenvectors can be constructed acting on the vacuum of the lowering operators with powers of the raising operators. 
However, in \cite{BGR}, we proved that their approach is only formal, meaning with this that the objects they work with are intrinsically ill-defined. In fact, we proved that the only vacuum of $H$ is a Dirac delta distribution. 

Recently, two of the three authors of \cite{nakano} produced a new paper to show that our main conclusion in \cite{BGR} is wrong. In fact, this is not the case, as it is quite easy to show. This is the content of Section \ref{sect3}, which follows a section with a short review of our results in \cite{BGR}. In section \ref{sect4} we  propose a very simple example useful to clarify what is going on.

\section{A short review}\label{sect2}

The classical equation for the DHO is $m\ddot x+\gamma \dot x+kx=0$, in which $m,\gamma$ and $k$ are the physical positive quantities of the oscillator: the mass, the friction coefficient and the spring constant. The Bateman lagrangian is
\be L=m\dot x\dot y+\frac{\gamma}{2}(x\dot y-\dot xy)-kxy,
\label{21}\en
which other than the previous equation, produces also $m\ddot y-\gamma \dot y+ky=0$, the differential equation associated to the AHO. Introducing the conjugate momenta  $p_x=\frac{\partial L}{\partial \dot x}=m\dot y-\frac{\gamma}{2}\,y,$ and
$p_y=\frac{\partial L}{\partial \dot y}=m\dot x+\frac{\gamma}{2}\,y,
$
the Hamiltonian looks as
\be
H=p_x\dot x+p_y \dot y-L=\frac{1}{m} p_xp_y+\frac{\gamma}{2m}\left(yp_y-xp_x\right)+\left(k-\frac{\gamma^2}{4m}\right)xy.
\label{22}\en
By introducing the new variables $x_1$ and $x_2$ through 
\be
x=\frac{1}{\sqrt{2}}(x_1+x_2), \qquad y=\frac{1}{\sqrt{2}}(x_1-x_2),
\label{23}\en
$L$ and $H$ can be written as follows:
$$
L=\frac{m}{2}(\dot x_1^2-\dot x_2^2)+\frac{\gamma}{2}(x_2\dot x_1-x_1\dot x_2)-\frac{k}{2}(x_1^2-x_2^2)
$$
and
$$
H=\frac{1}{2m}\left(p_1-\frac{\gamma}{2}x_2\right)^2-\frac{1}{2m}\left(p_2-\frac{\gamma}{2}x_1\right)^2+\frac{k}{2}(x_1^2-x_2^2),
$$
where $p_1=\cfrac{\partial L}{\partial \dot x_1}=m\dot x_1+\cfrac{\gamma}{2}\,x_2$ and $p_2=\cfrac{\partial L}{\partial \dot x_2}=m\dot x_2-\cfrac{\gamma}{2}\,x_1$. By putting $\omega^2=\cfrac{k}{m}\,-\cfrac{\gamma^2}{4m^2}$ we can further rewrite $H$ as follows:
\be
H=\left(\frac{1}{2m}p_1^2+\frac{1}{2}m\omega^2x_1^2\right)-\left(\frac{1}{2m}p_2^2+\frac{1}{2}m\omega^2x_2^2\right)-\frac{\gamma}{2m}(p_1x_2+p_2x_1).
\label{24}\en
To fix the ideas, we will restrict here to $\omega^2>0$.  

Following \cite{nakano} we impose the following canonical quantization rules between $x_j$ and $p_k$: $[x_j,p_k]=i\delta_{j,k}\mathds{1}$, working in unit $\hbar=1$. Here $\mathds{1}$ is the identity operator. This is equivalent to the choice in \cite{fesh}. Then we put
\be a_k=\sqrt{\frac{m\omega}{2}}\,x_k+i\sqrt{\frac{1}{2m\omega}}\,p_k,
\label{25}\en
$k=1,2$, which satisfy the canonical commutation rules: $[a_j,a^\dagger_k]=\delta_{j,k}\mathds{1}$. Hence we can write
\be
\left\{
\begin{array}{ll}
	H=H_0+H_I,\\
H_0=\omega\left(a_1^\dagger a_1-a_2^\dagger a_2\right),\\
	H_I=\cfrac{i\gamma}{2m}\left(a_1a_2-a_1^\dagger a_2^\dagger\right)\\
\end{array}
\right.
\label{26}\en 
which can be still be rewritten as
\be
\left\{
\begin{array}{ll}
	H=H_0+H_I,\\
	H_0=\omega\left(B_1A_1-B_2A_2\right),\\
	H_I=\cfrac{i\gamma}{2m}\left(B_1A_1+B_2A_2+\mathds{1}\right),\\
\end{array}
\right.
\label{210}\en 
which only depends on the pseudo-bosonic number operators $N_j=B_jA_j$, \cite{baginbagbook}. Here 
\be
A_1=\frac{1}{\sqrt{2}}(a_1-a_2^\dagger), \quad A_2=\frac{1}{\sqrt{2}}(-a_1^\dagger+a_2),
\label{27}\en
while
\be
B_1=\frac{1}{\sqrt{2}}(a_1^\dagger+a_2), \quad B_2=\frac{1}{\sqrt{2}}(a_1+a_2^\dagger).
\label{28}\en
These operators satisfy the following requirements:
\be
[A_j,B_k]=\delta_{j,k}\mathds{1},
\label{29}
\en
with $B_j\neq A_j^\dagger$, $j=1,2$.

 This is exactly the same Hamiltonian found in \cite{nakano}, and it is equivalent to the one given in \cite{dekk,fesh} and in many other papers on this subject. In \cite{nakano}, the authors introduce the vacuum for the annihilation operators $A_1$ and $A_2$ as the action of an unbounded operator on the vacuum of $a_1$ and $a_2$, and they construct new vectors out of this vacuum, claiming that these vectors, all together, form a Fock basis with norm equal to one. In \cite{BGR} we proved that this is wrong and, in \cite{deguchi}, the authors replied that their results are correct, and provided an explicit expression of square integrable vacuum for their annihilation operators. 
 
 The present note is intended to show that the vacuum computed in \cite{deguchi} is not annihiled by the lowering operators, how can be checked by using a direct computation. Then we explain the main weakness of their argument, discussing also a very simple example to clarify the situation. However, before doing this, we state our main proposition proven in \cite{BGR}, which shows that no square-integrable vacuum of the lowering operators can be found. We refer to \cite{BGR} for the simple proof of the result, for some useful comments, and for the analysis of the overdamped case, $\omega^2<0$. 

\begin{prop}\label{prop1}
	There is no non-zero function $\varphi_{00}(x_1,x_2)$ satisfying $$A_1\varphi_{00}(x_1,x_2)=A_2\varphi_{00}(x_1,x_2)=0.$$ Also, there is no non-zero function $\psi_{00}(x_1,x_2)$ satisfying $$B_1^\dagger\psi_{00}(x_1,x_2)=B_2^\dagger\psi_{00}(x_1,x_2)=0.$$
\end{prop}

\section{What is wrong with \cite{deguchi} and \cite{nakano}}\label{sect3}

We begin this section by showing, with a direct computation, that the paper in \cite{deguchi} contains an essential (and trivial) mistake, which makes all their results incorrect. To avoid possible misunderstanding, we adopt their notation, with the only difference that we keep $\hbar=1$ here.

In \cite{deguchi} the authors introduce $a_k$ as in  (\ref{25}) and use it to introduce new operators $\overline{a}_j$ and $\overline{a}_j^\ddagger$ as follows
$$
\overline{a}_1=\frac{1}{\sqrt{2}}\left(a_1\mp a_2^\dagger\right), \qquad \overline{a}_2=\frac{1}{\sqrt{2}}\left(\mp a_1^\dagger+ a_2\right),$$  and $$ \overline{a}_1^\ddagger=\frac{1}{\sqrt{2}}\left(a_1^\dagger\pm a_2\right), \qquad \overline{a}_2^\ddagger=\frac{1}{\sqrt{2}}\left(\pm a_1+ a_2^\dagger\right).
$$
Comparing these formulas with (\ref{27}) and (\ref{28}), we see that these are closely related to our operators $A_j$ and $B_j$. It may be worth noticing that these definitions should be clarified, due to the presence of the $\pm$ and $\mp$ in the formulas. For this reason, we prefer to call, for instance,
$$
\overline{a}_{1,-}=\frac{1}{\sqrt{2}}\left(a_1- a_2^\dagger\right)=\frac{1}{\sqrt{2}}\left[\sqrt{\frac{m\omega}{2}}(x_1-x_2)+\sqrt{\frac{1}{2m\omega}}\left(\frac{\partial}{\partial x_1}+\frac{\partial}{\partial x_2}\right)\right].
$$
Now, it is trivial to check that their {\em proposed vacuum}, see formula (27) in \cite{deguchi},
$$
\overline{\varphi}_{0,0}(x_1,x_2)=\sqrt{\frac{m\omega}{\pi}}\exp\left\{-\frac{m\omega}{2}(x_1^2+x_2^2)\right\},
$$
does not obey $\overline{a}_{1,-}\overline{\varphi}_{0,0}(x_1,x_2)=0$. In fact, we get $$\overline{a}_{1,-}\overline{\varphi}_{0,0}(x_1,x_2)=-\sqrt{m\omega}\,x_2\,\overline{\varphi}_{0,0}(x_1,x_2).$$  Similarly we can check that $\overline{a}_{2,-}\overline{\varphi}_{0,0}(x_1,x_2)\neq 0$. Hence, $\overline{\varphi}_{0,0}(x_1,x_2)$ is not the vacuum of $\overline{a}_{1,-}$, contrarily to what is claimed in \cite{deguchi}.

\subsection{More comments on \cite{deguchi} and \cite{nakano}}

However, in view of their relevance for a deeper understanding of what is going on, we would like to add some remarks on the mathematical weakness of the two papers \cite{deguchi} and \cite{nakano}. The first remark is related to the operator $e^{\theta X}$ they introduce, where $X=a_1a_2+a_1^\dagger a_2^\dagger$. They use $e^{\theta X}$ to define the vacuum of their $\overline{a}_j$ via its action on the vacuum of the $a_j$: if $a_j|0>=0$, $j=1,2$, their claim is that, calling $|0\rangle\!\rangle=e^{\theta X}|0>$, then $\overline{a}_j|0\rangle\!\rangle=0$, $j=1,2$. Their argument is based on the fact that $\overline{a}_j=e^{\theta X} a_j e^{-\theta X}$, for suitable choices of $\theta$, and on the formal result: $\overline{a}_j|0\rangle\!\rangle=e^{\theta X} a_j e^{-\theta X}e^{\theta X}|0>=e^{\theta X} a_j |0>=0$.
 However the authors do not consider the fact that  $e^{\theta X}$ is unbounded, and it can easily happen that $|0>$ does not belong to the domain of $e^{\theta X}$, $D(e^{\theta X})$, and therefore it is not granted that $e^{\theta X}|0>$ makes any sense. This will be clarified in the simple example given in the next section.

In formula (16) of \cite{deguchi} they consider the scalar product $\langle x_1,x_2|0\rangle\!\rangle$, calling this result $\varphi_{0,0}(x_1,x_2)$. The obvious problem is that, also in connection to our previous remark, $|0\rangle\!\rangle$ is not necessarily an element in the Hilbert space, and therefore the scalar product is not, most likely, well defined. This kind of problems are scattered everywhere in the paper, and not properly considered. The only attempt to clarify the situation is in the introduction of the sets $\cal B$ and $\cal K$ as the {\em bra and ket spaces for the $(a_i,a_i^\dagger)$-system, respectively}, and their counterparts $\overline{\cal B}$ and $\overline{\cal K}$. The point is that these sets are only vaguely introduced, while no mathematical detail is given at all (they talk of an {\em improper inner product}, without any clarification).

We end this list of comments by noticing a last serious mathematical inaccuracy. The operators $a_i$ and $a_i^\dagger$ are defined on the sets $\cal B$ and $\cal K$, the authors claim, (again, giving no mathematical definition for these sets). And they conclude that the vectors in (20) of \cite{deguchi} are elements of $\cal K$. It is not clear why this should be true, firstly because we should understand how $\cal K$ is defined to prove what they say. And secondly, since it is well known that unbounded operators can easily map a dense subspace of an Hilbert space into a different space. Hence it is not enough to know that $a_i$ and $a_i^\dagger$ are defined on  $\cal K$ to conclude that, say, $a_i^\dagger f\in\cal K$ for all $f\in\cal K$.

More comments could be given. However, we prefer to produce in the next section a simple example which clarifies that {\em strange things may happen}, when unbounded operators are involved in the analysis of some physical system.

\section{A pedagogical example}\label{sect4}

Let $x$ and $p$ be the position and momentum operators, $[x,p]=i\1$, and $c=\frac{1}{\sqrt{2}}(x+ip)$ the related annihilation bosonic operator. We know that $[c,c^\dagger]=\1$, and it is easy to find the vacuum of $c$, $c\varphi_0(x)=0$, since it must satisfy the differential equation $\varphi_0'(x)=-x\varphi_0(x)$. This is because $p=-i\frac{d}{dx}$. Hence $\varphi_0(x)=Ne^{-x^2/2}$, with $N$ a suitable normalization. It is clear that $\varphi_0(x)\in\Lc^2(\mathbb{R})$. It is also well known that no square-integrable vacuum exists for $c^\dagger=\frac{1}{\sqrt{2}}(x-ip)$, since the solution of $c\psi_0(x)=0$ is proportional to $e^{x^2/2}$. Of course, we could still try to work with $\psi_0(x)$ in some different Hilbert space, introducing some metric on $\Lc^2(\mathbb{R})$ and working with a different scalar product, in order to have $\psi_0(x)$ square integrable. But this would modify the notion of the adjoint, and therefore $\psi_0(x)$ needs not to be the vacuum of this new adjoint of $c$, $c^\sharp$. 

Going back to our original problem, let us consider the operator $T=e^{\frac{7\pi}{8}\left(c^2+(c^\dagger)^2\right)}$. This operator is unbounded, invertible, and (formally) self-adjoint. We want to show that working with $T$ as if it was a bounded operator creates, as in \cite{nakano} and \cite{deguchi}, paradoxes. Hence, from now on, we will work formally, paying no attention to domains of operators and see that something strange happens.

First of all, it is easy to check that
$$
x=\frac{1}{\sqrt{2}}(c+c^\dagger)=TcT^{-1}.
$$
Now, defining $\Phi_0(x)=T\varphi_0(x)$, we should have, similarly to what is done in \cite{deguchi} and \cite{nakano},
$$
x\,\Phi_0(x)=(TcT^{-1})(T\varphi_0(x))=T\,c\,\varphi_0(x)=0.
$$
Hence $\Phi_0(x)$ should satisfy $x\,\Phi_0(x)=0$. But the only function which solves this equation is $\Phi_0(x)=0$, which is not compatible with the existence of $T^{-1}$ and with the fact that $\varphi_0(x)\neq0$. Of course, a non trivial solution does exist, but only in a distributional sense: $\Phi_0(x)=N'\delta(x)$.

\vspace{2mm} {\bf Remark:--} The same conclusion can be deduced by noticing that, with a little algebra, 
$$
T\varphi_0= e^{-\frac{1}{2}{c^\dagger}^2}\,e^{\frac{1}{4}\log2(cc^\dagger+c^\dagger c)}\, e^{-\frac{1}{2}{c}^2}\varphi_0=2^{1/4}e^{-\frac{1}{2}{c^\dagger}^2}\varphi_0=2^{1/4}\sum_{k=0}^\infty\frac{\sqrt{2k}!}{k!}\,\left(-\frac{1}{2}\right)^k\,\varphi_{2k}.
$$
{The operator $T$ has been factorized by using standard operators ordering properties (see for instance \cite{rad}, Appendix 5), and  $\{\varphi_k\}_{k\geq0}$ is the basis of $\Lc^2(\mathbb{R})$  made by the  eigenstates of the quantum harmonic oscillator.}
Now, using the Raabe's test, it is possible to check that the series for $\|T\varphi_0\|^2$ diverges. Hence $T\varphi_0$ is not a vector in $\Lc^2(\mathbb{R})$, as we have explicitly shown above.

\vspace{2mm}

This simple example shows what is going on with the DHO, and should clarify the role of unbounded operators and of distributions in this context.

\section*{Acknowledgements}
The authors acknowledge partial support from Palermo University. F.B. and F.G.  acknowledge partial support from G.N.F.M. of the I.N.d.A.M.
F.G. acknowledges support by M.I.U.R.

\end{document}